# Superconductivity in Nb$_2$Pd$_3$Te$_5$ and Chemically-Doped Ta$_2$Pd$_3$Te$_5$

Naoya Higashihara[1], Yoshihiko Okamoto[1,*], Yuma Yoshikawa[1], Youichi Yamakawa[2], Hiroshi Takatsu[3], Hiroshi Kageyama[3], and Koshi Takenaka[1]

[1]*Department of Applied Physics, Nagoya University, Nagoya 464-8603, Japan*
[2]*Department of Physics, Nagoya University, Nagoya 464-8602, Japan*
[3]*Graduate School of Engineering, Kyoto University, Kyoto 615-8510, Japan*

We report on the superconductivity in ternary transition metal tellurides Ta$_2$Pd$_3$Te$_5$ and Nb$_2$Pd$_3$Te$_5$, which have a one-dimensional crystal structure. Single-crystalline and polycrystalline samples of Ta$_2$Pd$_3$Te$_5$ show nonmetallic electron conduction, while Ti or W doping results in metallic behavior with a bulk superconducting transition at 2-4 K. In contrast, the polycrystalline samples of Nb$_2$Pd$_3$Te$_5$, which are found to be isostructural to Ta$_2$Pd$_3$Te$_5$, show a bulk superconducting transition at 3.3 K. The crystal structure and physical properties of Ta$_2$Pd$_3$Te$_5$ and Nb$_2$Pd$_3$Te$_5$ are also discussed in comparison with a candidate excitonic insulator Ta$_2$NiSe$_5$.

Transition metal tellurides are known to show various outstanding *d*-electron properties, mainly due to their low-dimensional crystal structures caused by the chemical bonding of tellurium atoms. Type-II Weyl semimetals WTe$_2$ and TaIrTe$_4$ and one-dimensional Dirac semimetals Ta$_4$SiTe$_4$ and ZrTe$_5$ are recent examples.[1-5] There are also many unconventional superconductors in transition metal tellurides.[6-8] For example, a type-II Dirac semimetal PdTe$_2$ showed a type-I bulk superconductivity accompanied by a possible surface superconductivity related to its topological surface states.[6] All of the above tellurides are van der Waals crystals, where slub- or chain-like structural blocks are weakly bonded by van der Waals interactions working between tellurium atoms, which results in highly one- or two-dimensional crystal and electronic structures.

In this study, we focus on $A_2$Pd$_3$Te$_5$ ($A$ = Ta, Nb) as a new member of the superconducting transition metal tellurides. Ta$_2$Pd$_3$Te$_5$ was reported to crystallize in an orthorhombic crystal structure with a space group of *Pnma*.[9] As shown in Fig. 1(a), in Ta$_2$Pd$_3$Te$_5$, a Ta atom is square-pyramidally coordinated by five Te atoms, and the TaTe$_5$ pyramids form one-dimensional chains along the *b*-axis by sharing their edges. Pd atoms occupy interstitial sites between the chains, which are tetrahedrally coordinated by Te atoms. Thus, the formed Ta$_2$Pd$_3$Te$_5$ layers are stacked along the *a*-axis via van der Waals bonds between Te atoms (Fig. 1(b)). The electrical resistivity, ρ, of the single crystals showed nonmetallic behavior with $d\rho/dT$ < 0, and the angle-resolved photoemission and scanning tunneling spectra indicated the presence of a bulk band gap of several tens of meV at the Fermi energy and topological edge states in the band gap.[10,11] Other than Ta$_2$Pd$_3$Te$_5$, Ta$_4$Pd$_3$Te$_{16}$ and Ta$_3$Pd$_3$Te$_{14}$ have been reported as the members of the Ta-Pd-Te ternary system.[12,13]

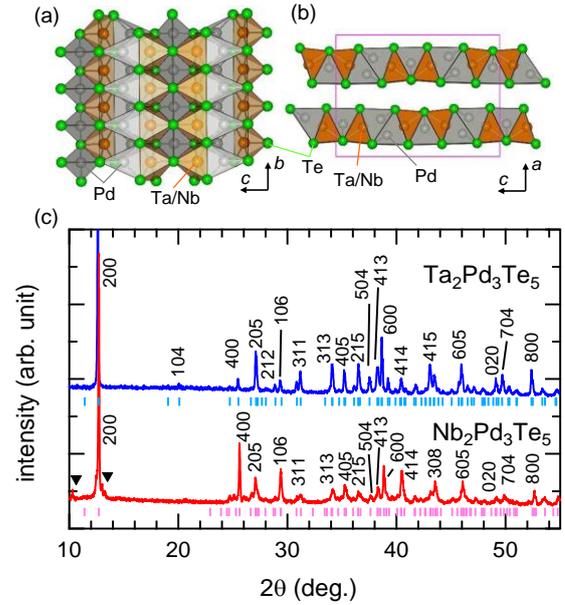

Fig. 1. (a, b) Crystal structures of $A_2$Pd$_3$Te$_5$ ($A$ = Ta, Nb) viewed along the *a*- and *b*-axes. The solid line in (b) represents the unit cell. (c) Powder XRD patterns of the Ta$_2$Pd$_3$Te$_5$ (upper) and Nb$_2$Pd$_3$Te$_5$ (lower) polycrystalline samples measured at room temperature. The peaks indicated by triangles are those of unknown impurities. The upper and lower vertical bars indicate the positions of the Bragg reflections of Ta$_2$Pd$_3$Te$_5$ and Nb$_2$Pd$_3$Te$_5$, respectively. The reflections that have an intensity greater than 0.5% of the strongest peak are shown.

Ta$_2$Pd$_3$Te$_5$ is interesting in that its crystal structure is similar to that of the candidate excitonic insulator Ta$_2$NiSe$_5$;[18,19] the complete removal of Pd atoms from two of the three Pd sites in the Ta$_2$Pd$_3$Te$_5$ layer results in an identical $Ta_2NiSe_5$ layer. In this regard, $Ta_2Pd_3Te_5$ has a "stuffed" $Ta_2NiSe_5$-type structure.[9] $Ta_2NiSe_5$ shows a second-order structural phase transition from orthorhombic to monoclinic symmetry at 326 K under ambient pressure, which is accompanied by anomalies in electrical resistivity and magnetic susceptibility.[20,21] This transition was understood as the formation of an excitonic insulator state with a finite exciton gap.[22] This transition temperature decreased by applying a pressure, and a superconducting transition at $T_c \sim$ 1 K appeared at around the critical point.[23] In contrast, $Nb_2Pd_3Te_5$, a $4d$ analogue of $Ta_2Pd_3Te_5$, has not yet been synthesized. In the Nb-Pd-Te ternary system, only $NbPdTe_5$ was synthesized, which was reported to show metallic $\rho$.[24,25]

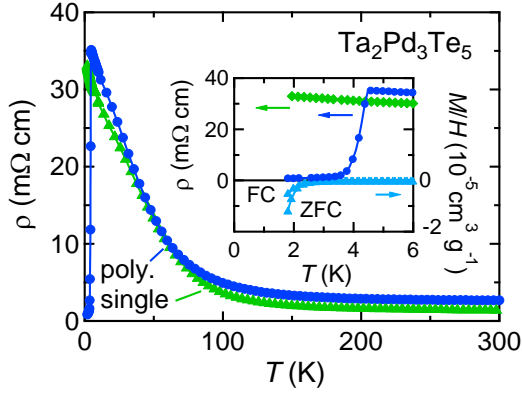

Fig. 2. Temperature dependence of electrical resistivity of the single-crystalline and polycrystalline samples of $Ta_2Pd_3Te_5$. For the single crystal data, electric current was applied parallel to the $b$-axis. The inset shows an enlarged view of the main panel and the temperature dependence of the zero-field-cooled and field-cooled magnetization of the $Ta_2Pd_3Te_5$ polycrystalline sample measured under a magnetic field of 20 Oe.

A series of Ti- and W-doped polycrystalline samples of $Ta_2Pd_3Te_5$ and $Nb_2Pd_3Te_5$ and the single crystals of $Ta_2Pd_3Te_5$ were prepared by the solid-state reaction and chemical vapor transport methods, respectively.[26] Sample characterization was performed by powder XRD analysis with Cu Kα radiation at room temperature using a MiniFlex diffractometer (Rigaku). As shown in Fig. 1(c), almost all diffraction peaks in the $Ta_2Pd_3Te_5$ data could be indexed on the basis of an orthorhombic structure with lattice constants of $a = 13.964(14)$ Å, $b = 3.704(5)$ Å, and $c = 18.62(6)$ Å, which are almost identical to those of the previous studies.[9,10] The powder XRD pattern of $Nb_2Pd_3Te_5$ is similar to that of $Ta_2Pd_3Te_5$, and almost all diffraction peaks could be indexed with lattice constants $a = 13.8994(11)$ Å, $b = 3.7097(13)$ Å, and $c = 18.686(6)$ Å. These lattice constants are almost identical to those of $Ta_2Pd_3Te_5$, and the extinction rule that appeared in the pattern was the same as that of $Ta_2Pd_3Te_5$, indicating that $Nb_2Pd_3Te_5$ is isostructural to $Ta_2Pd_3Te_5$. The diffraction patterns of Ti- or W-doped $A_2Pd_3Te_5$ samples do not show significant peaks due to impurity phases, indicating that the chemically doped samples were appropriately prepared.[27] The physical properties shown below are the results of $x \leq 0.3$ for $(A_{1-x}Ti_x)_2Pd_3Te_5$, $y \leq 0.3$ for $(Ta_{1-y}W_y)_2Pd_3Te_5$, and $y = 0.1$ for $(Nb_{1-y}W_y)_2Pd_3Te_5$. The electrical resistivity, magneti- zation, and heat capacity measurements were performed using a Physical Property Measurement System and Magnetic Property Measurement System (both Quantum Design). First principles calculations were performed using the WIEN2k code.[28] The crystal structure views were drawn using VESTA.[29]

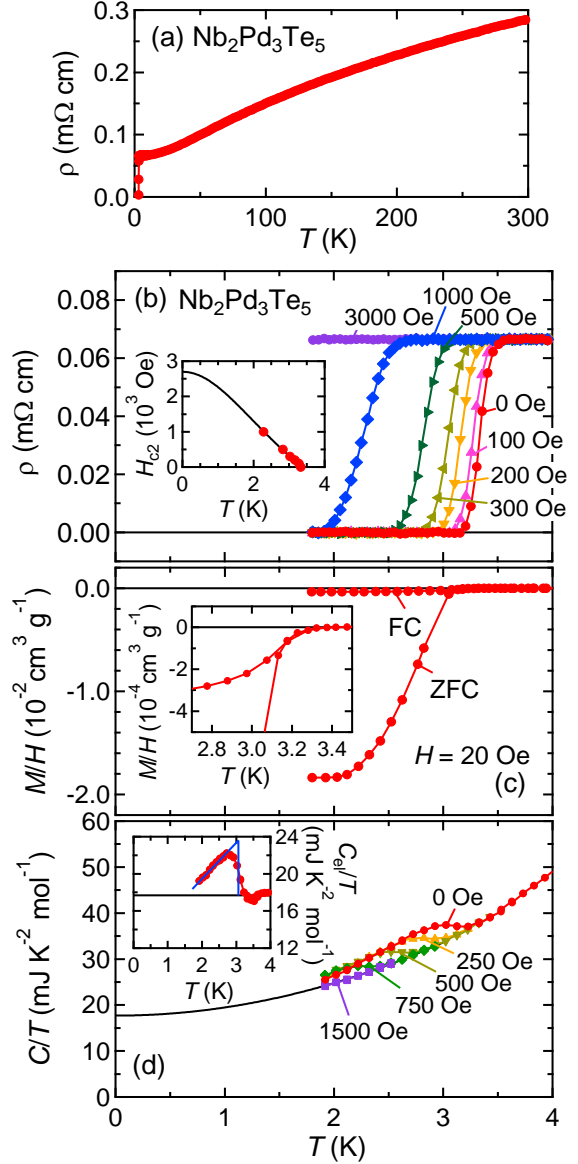

Fig. 3. Physical properties of the $Nb_2Pd_3Te_5$ polycrystalline samples. (a) Temperature dependence of the electrical resistivity. (b) Temperature dependences of the electrical resistivity measured under various magnetic fields of 0–3000 Oe. The inset shows the $H_{c2}$ determined by the midpoint of the resistivity drop. The solid curve shows a fit to the GL formula. (c) Temperature dependence of the zero-field-cooled and field-cooled magnetization measured under a magnetic field of 20 Oe. The inset shows an enlarged view of the main panel. (d) Temperature dependence of the heat capacity divided by temperature measured under various magnetic fields of 0–1500 Oe. The solid curve shows a fit of $C/T = AT^2 + \gamma$ to the normal-state data. The inset shows the electron heat capacity divided by temperature.

Figure 2 shows the temperature dependence of the ρ of single-crystalline and polycrystalline samples of $Ta_2Pd_3Te_5$. In the case of the single crystal, electric current was applied parallel to the $b$-axis. For both samples, the ρ values at 300 K were several mΩ cm and increased with decreasing temperature, reaching approximately 30 mΩ cm at 5 K. As shown in the inset of Fig. 2, there was no anomaly in ρ of the single crystal at low temperatures, but that of the polycrystalline sample sharply decreased from 4.5 K with decreasing temperature. This decrease shifted to lower temperature under application of a magnetic field, suggesting a superconducting transition. However, ρ did not become zero even at the lowest measured temperature of 1.8 K. The magnetization of the polycrystalline sample, shown in the inset of Fig. 2, exhibited a diamagnetic signal probably due to superconductivity, but the volume fraction was very small; the shielding fraction at 1.8 K was 0.08%. These results suggest that the superconductivity observed for the polycrystalline samples was extrinsic, probably due to impurities.

In contrast, the $Nb_2Pd_3Te_5$ polycrystalline samples showed bulk superconductivity. As shown in Figs. 3(a, b), the ρ showed a metallic behavior that gradually decreased with decreasing temperature, with a sharp drop to zero between 3.6 and 3.2 K. As shown in Fig. 3(c), the zero-field-cooled magnetization data showed a strong diamagnetic signal below 3.3 K, indicating that a bulk superconducting transition occurred at this temperature. The shielding fraction at 1.8 K was estimated to be a large value of 180%, well beyond 100%, probably due to the demagnetization effect.[30,31] The field-cooled data also showed a drop at 3.3 K and a small but finite Meissner signal below this temperature, which is typical behavior of a type-II superconductor with pinning.[32,33] The jump at around 3 K in the heat capacity data shown in Fig. 3(d) also supports the emergence of a bulk superconducting transition at this temperature. Considering the midpoint of the resistivity drop, the onset of the magnetization drop, and the onset of the heat capacity jump were at 3.3, 3.3, and 3.2 K, respectively, the superconducting transition temperature was determined to be 3.3 K.

We will now discuss the superconducting properties of $Nb_2Pd_3Te_5$. The inset of Fig. 3(b) shows the upper critical fields, $H_{c2}$, determined by the midpoint of the resistivity drop shown in the main panel. By fitting the Ginzburg-Landau (GL) formula $H_{c2}(T) = H_{c2}(0)[1 - (T/T_c)^2]/[1 + (T/T_c)^2]$ to the $H_{c2}$ data, $H_{c2}(0)$ and the GL coherence length $\xi_{GL}$ were estimated to be $2.70(9) \times 10^3$ Oe and 34.9(6) nm, respectively. The inset of Fig. 3(d) shows the electron heat capacity divided by temperature obtained by subtracting the lattice contribu-tion from the experimental data, $C_{el}/T = C/T - AT^2$. The lattice contribution, $AT^2$, was estimated by fitting the normal-state data below 3 K to $C/T = AT^2 + \gamma$, yielding $A$ = 1.77(2) mJ K$^{-4}$ mol$^{-1}$ and a Sommerfeld coefficient of $\gamma$ = 17.69(13) mJ K$^{-2}$ mol$^{-1}$. As shown in the inset of Fig. 3(d), the magnitude of the jump at $T_c$ was estimated to be $\Delta C_{el}/T_c$ = 6 mJ K$^{-2}$ mol$^{-1}$ by fitting the $C_{el}/T$ data with straight lines. These $\Delta C_{el}/T_c$ and $\gamma$ values yielded $\Delta C_{el}/\gamma T_c$ = 0.3, which is much smaller than the weak-coupling limit value of 1.43 for a conventional Bardeen–Cooper–Schrieffer superconductor. This result suggests that there might be a considerable contribution of conduction electrons that remained in the normal state down to the lowest temperature. It is unknown whether this contribution was due to the $Nb_2Pd_3Te_5$ phase or small amounts of impurity phases. To discuss the superconducting properties based on the heat capacity data in more detail, it might be necessary to reduce this contribution further.

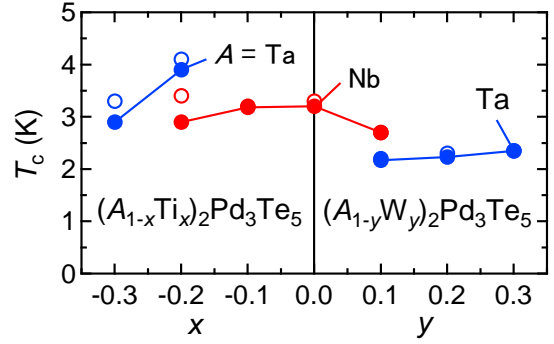

Fig. 4. The superconducting transition temperatures of $(A_{1-x}Ti_x)_2Pd_3Te_5$ and $(A_{1-y}W_y)_2Pd_3Te_5$ ($x$ = 0.1, 0.2, and 0.3 for $A$ = Ta and Nb, while $y$ = 0.1, 0.2, and 0.3 for $A$ = Ta and $y$ = 0.1 for $A$ = Nb) polycrystalline samples as a function of $x$ and $y$. The filled and open circles represent the midpoint of the resistivity drop and the onset of the magnetization drop, respectively.[34] The data for samples showing both zero resistivity and a large shielding fraction of several tens of percent or more are shown.

Thus, the ground states of undoped $Ta_2Pd_3Te_5$ and $Nb_2Pd_3Te_5$ were different, but in $Ta_2Pd_3Te_5$, bulk superconductivity appeared by chemical doping.[34] As shown in Fig. 4, undoped $Ta_2Pd_3Te_5$ did not show a bulk $T_c$ above 1.8 K, but the bulk $T_c$ at 2–4 K appeared in the Ti- and W-doped samples. Ti and W doping of the Ta sites simply correspond to hole and electron doping, respectively,[3,35] suggesting that the bulk superconductivity was induced by carrier doping. However, the effects of chemical pressure due to Ti and W doping and the Te deficiency induced by the presence of Ti and W atoms might also have an effect.[36] In contrast, $Nb_2Pd_3Te_5$ showed the highest $T_c$ of 3.3 K for the undoped sample, which was slightly decreased by Ti and W doping.[34]

The electronic band structure and electronic density of states (DOS) of the undoped $Ta_2Pd_3Te_5$ are shown in Fig. 5. The band dispersions along the $b^*$-axis (Γ–Y and R–U) were much stronger than those perpendicular to the $b^*$-axis, indicating the strongly one-dimensional electronic structure. Just above the Fermi energy, $E_F$, there was a valley of DOS with an almost constant and small value of approximately 5 eV$^{-1}$, yielding a calculated Sommerfeld coefficient of $\gamma_{band} = 3$ mJ K$^{-2}$ mol$^{-1}$, over an energy range of about 0.2 eV. The experimental $\gamma$ of the undoped $Ta_2Pd_3Te_5$ was 2.3 mJ K$^{-2}$ mol$^{-1}$, which is small and comparable to $\gamma_{band}$, suggesting that the undoped polycrystalline samples in this study were slightly electron doped for some reason, perhaps due to Te deficiency, and their $E_F$ was located in this valley. In this case, assuming a rigid-band structure, electron and hole dopings by substituting 1% of the Ta sites with Ti and W atoms, respectively, shifted the $E_F$ by $\pm 0.016$ eV. As a result of 10–20% Ti or W doping, the $E_F$ might be shift out of the DOS valley, giving rise to the metallic electrical conduction and bulk superconducting transition. In contrast, the undoped $Nb_2Pd_3Te_5$ showed metallic behavior with a bulk $T_c$ of 3.3 K. As shown in Fig. 3(d), the $\gamma$ of the undoped $Nb_2Pd_3Te_5$ was 17.7 mJ K$^{-2}$ mol$^{-1}$, which is about eight times larger than that of the undoped $Ta_2Pd_3Te_5$, suggesting that the DOS at the $E_F$ of $Nb_2Pd_3Te_5$ is much larger than that of $Ta_2Pd_3Te_5$. This difference is interesting in that the electronic structure of $Nb_2Pd_3Te_5$ near the $E_F$ might be significantly different from that of $Ta_2Pd_3Te_5$, despite the same electron configuration and crystal structure.

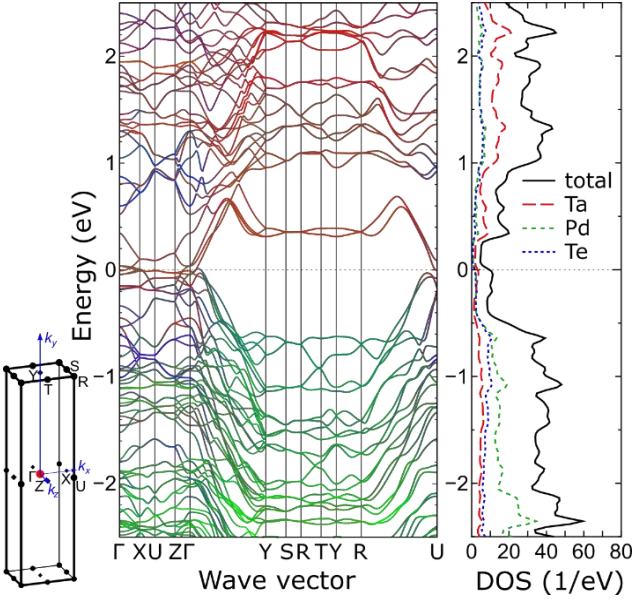

Fig. 5. The electronic states of $Ta_2Pd_3Te_5$ without spin-orbit coupling. The electronic band structure (left) and total and partial electronic DOS (right) are shown. The Fermi level is set to 0 eV. The first Brillouin zone is shown in the lower left.

The electronic structure of $Ta_2Pd_3Te_5$ shown in Fig. 5 is similar to that of $Ta_2NiSe_5$ at ambient pressure.[37,38] The conduction and valence bands of $Ta_2Pd_3Te_5$ and $Ta_2NiSe_5$ mainly come from Ta 5$d$ and Pd/Ni 3$d$/4$d$ orbitals, respectively, hybridized with Te/Se 5$d$/4$p$ orbitals. They have strong dispersions parallel to the one-dimensional chains. As discussed above, the $E_F$ of $Ta_2Pd_3Te_5$ might be located in the DOS valley, while $Nb_2Pd_3Te_5$ had a much larger DOS at the $E_F$ than that of $Ta_2Pd_3Te_5$ and showed a bulk superconductivity. On the other hand, $Ta_2NiSe_5$ is a semiconductor with a small energy gap at an ambient pressure, which changed to metallic with a superconducting transition at $T_c \sim$ 1 K by applying a pressure.[23,37] This analogy between $A_2Pd_3Te_5$ and $Ta_2NiSe_5$ might imply that the undoped $Ta_2Pd_3Te_5$ and $Nb_2Pd_3Te_5$ correspond to the ambient- and high-pressure phases of $Ta_2NiSe_5$, respectively. $Ta_2NiSe_5$ showed a superconducting transition only in the high-pressure phase, which limited the experiments that can be performed. In contrast, superconductivity in $Nb_2Pd_3Te_5$ appeared at ambient pressure, which can be accessed by various probes. It is expected that various experiments on $A_2Pd_3Te_5$ will not only uncover the physical properties of $A_2Pd_3Te_5$, but also help elucidate the relationship between superconductivity and an excitonic insulating state.

In summary, $Nb_2Pd_3Te_5$ was found to show a bulk superconducting transition at $T_c = 3.3$ K. In contrast, the $Ta_2Pd_3Te_5$ samples showed nonmetallic behavior without a bulk superconductivity above 1.8 K. The electrical resistivity of $Ta_2Pd_3Te_5$ changed to metallic by Ti- and W-doping at the Ta sites, resulting in bulk superconducting transitions at $T_c = $ 2–4 K. Since $Nb_2Pd_3Te_5$ and $Ta_2Pd_3Te_5$ have the same crystal structure and electronic configuration, it is interesting that they showed such contrasting electronic properties. In addition, their crystal and electronic structures and physical properties have much in common with a candidate excitonic insulator $Ta_2NiSe_5$. We hope that $A_2Pd_3Te_5$ will be a platform for studying the physics of superconductivity related to the excitonic insulators.

## ACKNOWLEDGMENTS

The authors are grateful to D. Hirai and Z. Hiroi for their support in heat capacity measurements and T. Yamauchi for his help in magnetization measurements. This work was partly carried out at the Materials Design and Characterization Laboratory under the Visiting Researcher Program of the Institute for Solid State Physics, University of Tokyo and supported by JSPS KAKENHI (Grant Numbers: 18H04314, 19H05823, 19K21846, and 20H02603) and the Asahi Glass Foundation.

*e-mail: yokamoto@nuap.nagoya-u.ac.jp

**Supplementary Note 1. Sample preparation and powder X-ray diffraction data of Ti- or W-doped $A_2Pd_3Te_5$.**

A series of Ti- and W-doped polycrystalline samples of $Ta_2Pd_3Te_5$ and $Nb_2Pd_3Te_5$ were prepared by the solid-state reaction method. A stoichiometric amount of elemental powders was mixed and sealed in an evacuated quartz tube. The tube was heated to and maintained at 873 K for 12 h, 1123 K for 48 h, and then furnace cooled to room temperature. The single crystals of $Ta_2Pd_3Te_5$, as shown in Supplementary Fig. 1, were prepared by the chemical vapor transport method. A mixture of 4 : 6 : 9 molar ratio of Ta, Pd, and Te was sealed in an evacuated quartz tube with iodine as a transport agent. The hot and cold sides of the tube were heated to and maintained at 1173 K and 1123 K, respectively, for 96 h, after the tube was kept at 773 K for 24 h. The mixture was put on the hot side, and single crystals 1–2 mm long were grown at the cold side.

Supplementary Fig. 2(a) shows powder X-ray diffraction (XRD) patterns of the $(Ta_{1-x}Ti_x)_2Pd_3Te_5$ and $(Ta_{1-y}W_y)_2Pd_3Te_5$ samples. All diffraction peaks observed in the patterns, with the exception of some small peaks caused by small amounts of unknown impurities, can be indexed on the basis of an orthorhombic cell, as in the case of $Ta_2Pd_3Te_5$. As shown in Supplementary Figs. 2(b, c), the 200 and 800 reflection peak systematically shifted to a higher angle with increasing the Ti content, $x$, or the W content, $y$. These results indicate that these Ti- or W-doped $Ta_2Pd_3Te_5$ samples were appropriately synthesized. Supplementary Fig. 3(a) shows powder XRD patterns of the $(Nb_{1-x}Ti_x)_2Pd_3Te_5$ and $(Nb_{0.9}W_{0.1})_2Pd_3Te_5$ polycrystalline samples. As in the cases of chemically-doped $Ta_2Pd_3Te_5$, almost all of the diffraction peaks can be indexed on the basis of an orthorhombic cell. As shown in Supplementary Figs. 3(b, c), the 200 and 600 reflection peak systematically shifted to a higher angle with increasing $x$ or $y$. These results indicated that these Ti- or W-doped $Nb_2Pd_3Te_5$ samples were appropriately synthesized.

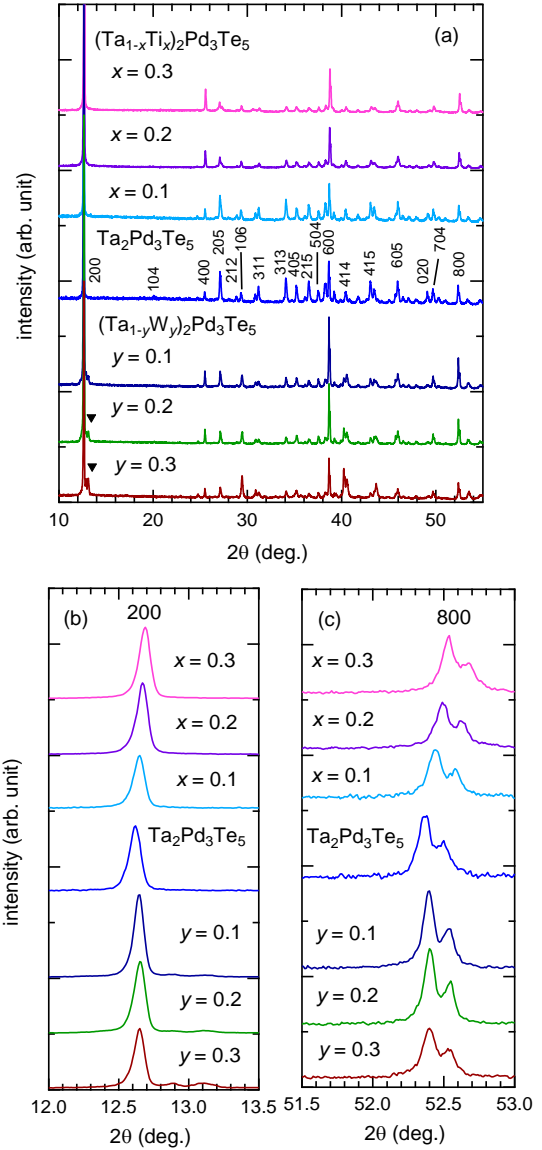

Supplementary Fig. 2. Powder XRD patterns (a) and those around the 200 (b) and 800 (c) reflection peaks of the $(Ta_{1-x}Ti_x)_2Pd_3Te_5$ ($x$ = 0.1, 0.2, and 0.3) and $(Ta_{1-y}W_y)_2Pd_3Te_5$ ($y$ = 0.1, 0.2, and 0.3) polycrystalline samples measured at room temperature. The data of undoped $Ta_2Pd_3Te_5$ presented in Fig. 1(c) are also shown as a reference. The peaks indicated by triangles are those of unknown impurities.

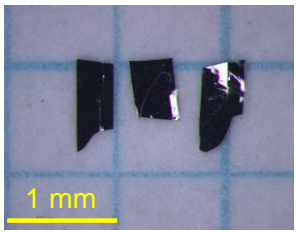

Supplementary Fig. 1. Single crystals of $Ta_2Pd_3Te_5$.

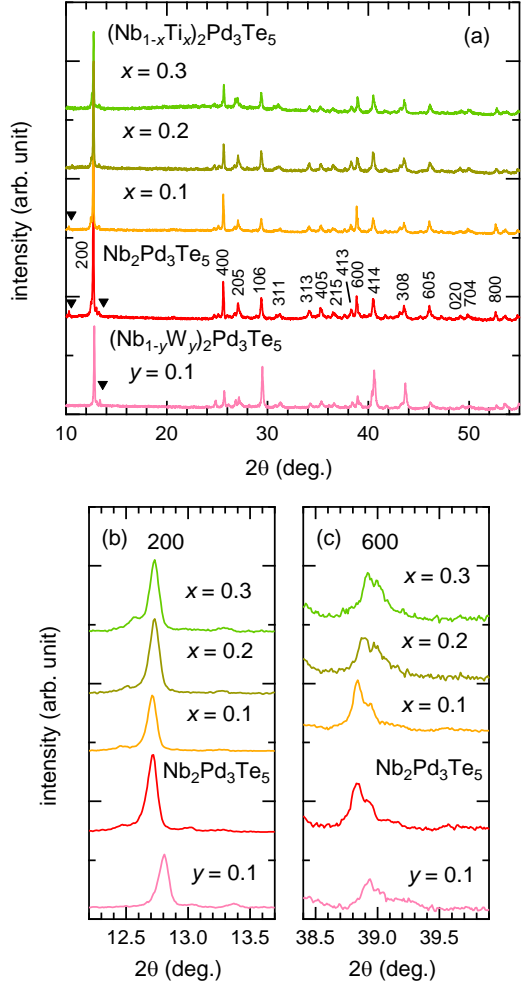

Supplementary Fig. 3. Powder XRD patterns (a) and those around the 200 (b) and 600 (c) reflection peaks of the $(Nb_{1-x}Ti_x)_2Pd_3Te_5$ ($x$ = 0.1, 0.2, and 0.3) and $(Nb_{1-y}W_y)_2Pd_3Te_5$ ($y$ = 0.1) polycrystalline samples measured at room temperature. The data of the undoped $Nb_2Pd_3Te_5$ presented in Fig. 1(c) are also shown as a reference. The peak indicated by the triangle is that of an unknown impurity.

## Supplementary Note 2. Physical Properties of Ti- or W-doped $A_2Pd_3Te_5$.

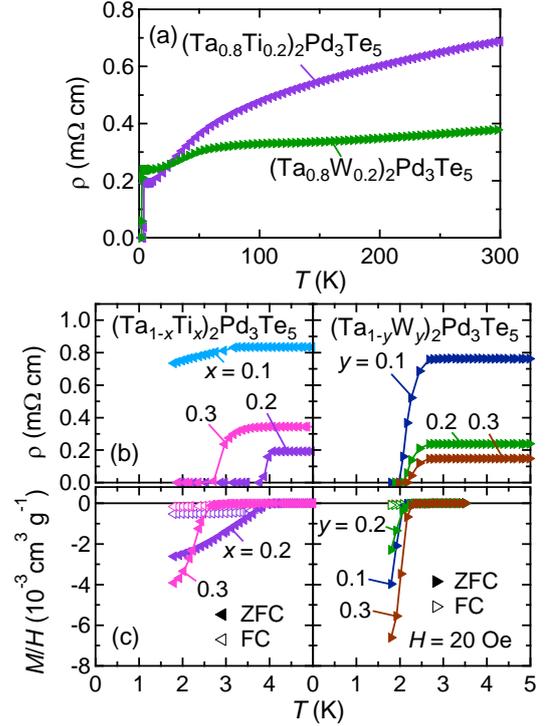

Supplementary Fig. 4. (a) Temperature dependence of the electrical resistivity of the $(Ta_{0.8}Ti_{0.2})_2Pd_3Te_5$ and $(Ta_{0.8}W_{0.2})_2Pd_3Te_5$ polycrystalline samples. (b,c) Temperature dependences of the electrical resistivity and zero-field-cooled and field-cooled magnetization of the $(Ta_{1-x}Ti_x)_2Pd_3Te_5$ (left) and $(Ta_{1-y}W_y)_2Pd_3Te_5$ (right) polycrystalline samples, respectively. Magnetization was measured under a magnetic field of 20 Oe.

Supplementary Figs. 4(a, b) show the temperature dependence of the electrical resistivity, ρ, of polycrystalline samples of $(Ta_{1-x}Ti_x)_2Pd_3Te_5$ and $(Ta_{1-y}W_y)_2Pd_3Te_5$ ($x, y$ = 0.1, 0.2, 0.3). The ρ of $Ta_2Pd_3Te_5$ was significantly reduced and changed to metallic by Ti or W doping. As shown in Supplementary Fig. 4(b), the $x \geq 0.2$ and $y \geq 0.1$ samples of $(Ta_{1-x}Ti_x)_2Pd_3Te_5$ and $(Ta_{1-y}W_y)_2Pd_3Te_5$, respectively, showed zero resistivity above 1.8 K. These samples also showed a large diamagnetic signal in the zero-field-cooled magnetization data shown in Supplementary Fig. 4(c). Their shielding fractions estimated with the 1.8 K data were several tens of percents, indicating that bulk superconductivity appeared for both of them. In addition, as shown in Supplementary Fig. 5, the $C/T$ data of the $y$ = 0.1 sample showed a clear anomaly supportive of bulk superconductivity.

Supplementary Figs. 6(a, b) show the temperature dependence of the electrical resistivity, ρ, of polycrystalline samples of $(Nb_{1-x}Ti_x)_2Pd_3Te_5$ ($x$ = 0.1, 0.2, and 0.3) and $(Nb_{0.9}W_{0.1})_2Pd_3Te_5$. All these samples showed metallic behavior with $d\rho/dT > 0$ and a zero resistivity at 2−3 K. The zero-resistivity temperatures of the Ti- or W-doped samples

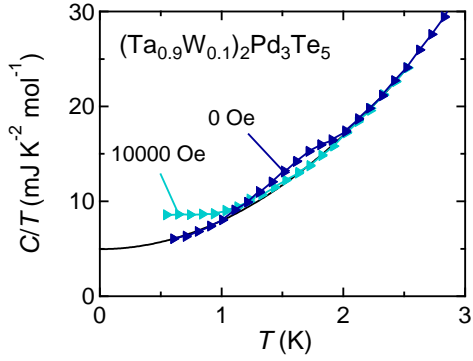

Supplementary Fig. 5. Temperature dependence of the heat capacity divided by the temperature of the $(Ta_{0.9}W_{0.1})_2Pd_3Te_5$ polycrystalline sample measured under a zero magnetic field and 10000 Oe. The solid curve shows a fit of the equation $C/T = AT^2 + \gamma$, yielding $A = 3.00(3)$ mJ K$^{-4}$ mol$^{-1}$ and $\gamma = 4.96(13)$ mJ K$^{-2}$ mol$^{-1}$.

and W-doped $Nb_2Pd_3Te_5$ samples showed a diamagnetic signal due to superconductivity. For the $x = 0.1$ and 0.2 of $(Nb_{1-x}Ti_x)_2Pd_3Te_5$ and $(Nb_{0.9}W_{0.1})_2Pd_3Te_5$ samples, the shielding fraction at 1.8 K was estimated to be several tens of percent, suggesting that the bulk superconducting transition occurred in them. In contrast, the shielding fraction in the $x = 0.3$ sample was a small value of 3%.

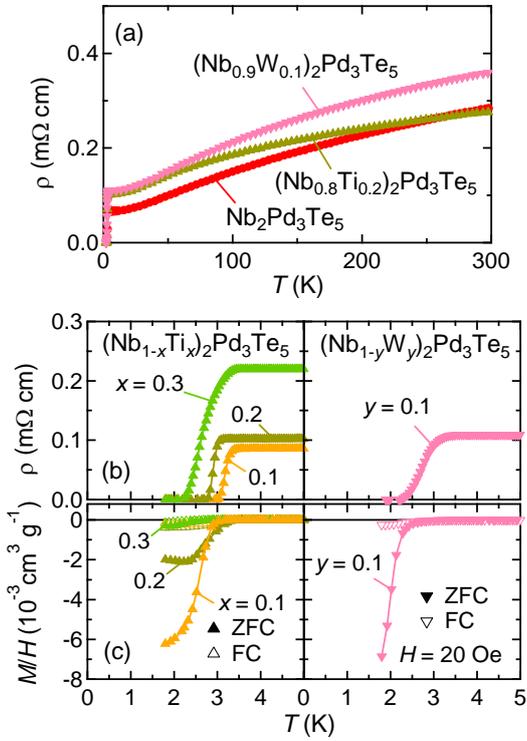

Supplementary Fig. 6. (a) Temperature dependence of the electrical resistivity of the $(Nb_{0.8}Ti_{0.2})_2Pd_3Te_5$, $(Nb_{0.9}W_{0.1})_2Pd_3Te_5$, and $Nb_2Pd_3Te_5$ polycrystalline samples. (b, c) Temperature dependences of the electrical resistivity and zero-field-cooled and field-cooled magnetization of the $(Nb_{1-x}Ti_x)_2Pd_3Te_5$ ($x = 0.1$, 0.2, and 0.3) (left) and $(Nb_{0.9}W_{0.1})_2Pd_3Te_5$ polycrystalline samples, respectively. Magnetization was measured under a magnetic field of 20 Oe.

were slightly lower than 3.2 K for the undoped $Nb_2Pd_3Te_5$. Furthermore, as shown in the left panel of Supplementary Fig. 6(b), the zero-resistivity temperature decreased with increasing Ti content $x$. As shown in the zero-field-cooled magnetization data shown in Supplementary Fig. 6(c), the Ti-